\documentclass[aps,pra,reprint,superscriptaddress,showpacs]{revtex4-1}
\usepackage{float}
\usepackage{amsfonts}
\usepackage{amsmath}
\usepackage{amssymb}
\usepackage{graphicx}
\usepackage{listings}
\usepackage{bm}
\usepackage{bbold}
\usepackage{braket}
\usepackage{todonotes}
\usepackage{hyperref}
\usepackage{array}
\usepackage{multirow}
\usepackage{url}
\usepackage[utf8]{inputenc}

 \newcolumntype{C}[1]{>{\centering\let\newline\\\arraybackslash\hspace{0pt}}m{#1}}

\newcommand{\dd}{\mathrm{d}}
\newcommand{\Tr}{\mathrm{Tr}}
\newcommand{\TE}{\mathrm{TE}}
\newcommand{\TM}{\mathrm{TM}}
\newcommand{\bulk}{\mathrm{bulk}}
\newcommand{\slab}{\mathrm{slab}}
\newcommand{\bk}{\mathbf{k}}
\newcommand{\Kappa}{{K}}
\newcommand{\Hb}{\overline{\mathrm{H}}}
\newcommand{\calR}{\mathcal{R}}
\newcommand{\calS}{\mathcal{S}}
\newcommand{\rmA}{\mathrm{A}}
\newcommand{\rmP}{\mathrm{P}}
\newcommand{\tpsi}{\widetilde{\psi}}
\newcommand{\tp}{\widetilde{p}}
\newcommand{\Hart}{\mathrm{E}_\mathrm{h}}
\newcommand{\Bohr}{\mathrm{a}_0}

\begin{document}

\title{Quantum reflection of antihydrogen from the Casimir potential
above matter slabs}

\author{G. Dufour}
\affiliation{Laboratoire Kastler-Brossel, CNRS, ENS, UPMC, Campus
Jussieu, F-75252 Paris, France}
\author{A. G{\'e}rardin}
\affiliation{Laboratoire Kastler-Brossel, CNRS, ENS, UPMC, Campus
Jussieu, F-75252 Paris, France}
\author{R. Gu{\'e}rout}
\affiliation{Laboratoire Kastler-Brossel, CNRS, ENS, UPMC, Campus
Jussieu, F-75252 Paris, France}
\author{A. Lambrecht}
\affiliation{Laboratoire Kastler-Brossel, CNRS, ENS, UPMC, Campus
Jussieu, F-75252 Paris, France}
\author{V.V. Nesvizhevsky}
\affiliation{Institut Laue-Langevin (ILL), 6 rue Jules Horowitz,
 F-38042, Grenoble, France}
\author{S. Reynaud}
\affiliation{Laboratoire Kastler-Brossel, CNRS, ENS, UPMC, Campus
Jussieu, F-75252 Paris, France}
\author{A.Yu. Voronin}
\affiliation{ P.N. Lebedev Physical Institute, 53 Leninsky prospect,
Ru-117924 Moscow, Russia}

\begin{abstract}
We study quantum reflection of antihydrogen atoms from matter slabs
due to the van der Waals/Casimir-Polder (vdW/CP) potential.  By
taking into account the specificities of antihydrogen and the
optical properties and width of the slabs we calculate realistic
estimates for the potential and quantum reflection amplitudes. Next
we discuss the paradoxical result of larger reflection coefficients
estimated for weaker potentials in terms of the Schwarzian
derivative. We analyze the limiting case of reflections at small
energies, which are characterized by a scattering length and have
interesting applications for trapping and guiding antihydrogen using
material walls.
\end{abstract}

 \maketitle

\section{Introduction}

Quantum reflection is the process of reflecting particles from an
attractive but rapidly varying potential. It has been studied since
the early days of quantum theory
\cite{LennardJones1936,*LennardJones1936a}. On the other hand it is
well known that atoms in the vicinity of a surface experience the
long-range van der Waals/Casimir-Polder potential (vdW/CP)
\cite{Casimir1946,*Casimir1948}. Quantum reflection occurs here if
the atom enters a region where the potential varies rapidly compared
with the atom's wavelength. Experimentally quantum reflection on the
vdW/CP potential has been observed with slow atoms reflected from a
liquid Helium surface \cite{NayakPRL1983,BerkhoutPRL1989,Yu1993} or
from solid surfaces \cite{Shimizu2001,Druzhinina2003}. More recent
efforts have focused on quantum reflection from rough or
micro-/nanostructured surfaces
\cite{Oberst2005,CroninPRA2009,Zhao2010} and on quantum reflection
of Bose-Einstein condensates on flat or nanostructured silicon
\cite{Pasquini2004,Pasquini2006}.

The theoretical description of quantum reflection has been the topic
of numerous contributions in the past
\cite{Berry1972,BrenigZeitPhysB1980,Clougherty1992,Henkel1996,Carraro1998,
Friedrich2002}, which are presented in some detail in
\cite{Friedrich2004}. The particular case of reflection on a vdW/CP
potential created by thin slabs or graphene sheets has been studied
recently in \cite{Judd2011}. In addition it has been put forward
that quantum reflection coefficients can be tuned using external
optical fields \cite{CroninPRA2008} or via thermal non-equilibrium
effects \cite{DruzhininaPRA2010}.

In the present paper, we will study the quantum reflection of
antihydrogen atoms $\Hb$ falling on material walls. As $\Hb$ atoms
are annihilated in contact with matter, this case enforces specific
boundary conditions at the material surface
\cite{Voronin2005,Voronin2005a}. In particular the behavior of the
short-range atom-wall potential becomes irrelevant as all antiatoms
that come close enough to the surface are annihilated. This topic is
important to the GBAR collaboration which aims to measure the
gravitational behavior of $\Hb$ by studying its time of free fall
from a well-defined trap to a matter plate \footnote{Proposal to
Measure the Gravitational Behavior of Antihydrogen at Rest (P.
P{\'e}rez \textit{et al.}, 2011); see
\url{http://gbar.in2p3.fr/public/SPSC-111025.pdf}}. We will give
accurate estimations for the van der Waals/Casimir-Polder (vdW/CP)
potential between the antiatoms and the surface as well as for the
associated quantum reflection.

A number of different methods are available to calculate
atom-surface dispersion forces
\cite{Feinberg1970,BabikerJPA1976,WyliePRA1984,BabbPRA2004} (see
\cite{Milton2011} for a detailed bibliography). Here we will use the
scattering approach \cite{Lambrecht2006,Emig2007} which has been
developed to calculate Casimir forces in arbitrary geometries and
which can be applied to the study of vdW/CP forces between an atom
and flat or nanostructured surfaces \cite{Messina2009}. In order to
obtain accurate estimations, it will in particular be necessary to
take into account the material properties and the finite thickness
of the slabs \cite{Lambrecht2007}.

In order to explain the paradoxical result that larger reflections
are obtained for weaker potentials, we will discuss how the quantum
reflection occurs when the atoms approach the surface and draw a
relation to the Schwarzian derivative. We will finally analyze the
limiting case of reflections at small energies, which have
interesting applications for trapping and guiding antihydrogen with
material walls \cite{Voronin2011,Voronin2012,*Voronin2012a}. Quantum
reflection is characterized by a scattering length which we will
calculate for different materials and different slab widths. We note
at this point that quantum reflection is calculated in the present
paper from a static potential, so that the role of dissipation in
matter is neglected \cite{Zhang2012}.

\section{Casimir-Polder potential}

We use the scattering formalism \cite{Lambrecht2006,Emig2007}
applied here to the Casimir-Polder potential between an atom and a
plate~:
\begin{equation}
 V(z)=\hbar \int_0^\infty \frac{\dd\xi}{2\pi} \Tr
 \ln \left( 1-\calR_\rmP e^{-\kappa z} \calR_\rmA e^{-\kappa
 z}\right)~.\label{general}
\end{equation}
As the quantum reflection process is expected to occur at distances
smaller than $1\mu$m (more discussions below), and thus smaller than
the typical thermal wavelength, this formula has been written at
zero temperature. The matrices $\calR_\rmP$ and $\calR_\rmA$
describe the reflection of the electromagnetic vacuum fields on the
plate and atom respectively. They are calculated for a Wick rotated
complex frequency $\omega=i\xi$ with the trace ($\Tr$) bearing on
transverse wave vectors $\bk$ and polarizations $p=\TE,\TM$. The
factor $e^{-\kappa z}$ accounts for propagation between the atom and
plate where $\kappa=\sqrt{\bk^2+\xi^2/c^2}$ is the Wick rotated
complex longitudinal wavevector.

We may safely neglect all multiple reflections between the atom and
the surface and thus expand the general scattering formula
(\ref{general}) to first order in $\calR_\rmA$. When the scattering
on the atom is described in the dipolar approximation
\cite{Messina2009}, the potential is read in terms of a dynamic
atomic polarizability $\alpha$, given in units of a volume:
\begin{align}
\label{vcp}
V(z)=&\frac{\hbar}{c^2} \int_0^\infty \; \dd\xi \xi^2
\alpha(i\xi)\;
\int \frac{\dd^2\bk}{(2\pi)^2} \frac{e^{-2 \kappa z}}{\kappa} \notag \\
&\times \left[ \rho^\TE-\left(1+\frac{2c^2k^2}{\xi^2}\right)\rho^\TM
\right]~.
\end{align}
The $\rho^p$ denote the electromagnetic reflection amplitudes for
the two polarizations $p=\TE,\TM$. We study first the case of
reflection from a semi-infinite bulk, described by the Fresnel laws
expressing continuity relations at the interface~:
\begin{equation}
\label{fresnel} \rho^\TE_\bulk= \frac{\kappa - \Kappa}{\kappa +
\Kappa}\quad, \quad \rho^\TM_\bulk=\frac{\epsilon(i\xi) \kappa -
\Kappa} {\epsilon(i\xi)\kappa + \Kappa}~,
\end{equation}
where $\Kappa=\sqrt{\bk^2+\epsilon(i\xi)(\xi/c)^2}$ corresponds to
the Wick rotated longitudinal wavevector inside the medium, and
$\epsilon$ is the relative dielectric function of this medium
(evaluated at the Wick rotated complex frequency).

The results presented below use the following optical response
properties~:
\begin{enumerate}
\item The atomic polarizability is that of antihydrogen ($\Hb$), and
is assumed to be the same as that of hydrogen (H)
\cite{Marinescu1997}.
\item Perfect mirrors have been used in previous calculations
\cite{Marinescu1997,Friedrich2002,Voronin2005,*Voronin2005a}
\begin{equation}
\label{perfect} \rho^\TE\equiv -1 \quad , \quad \rho^\TM\equiv 1~;
\end{equation}
they are considered here for the sake of comparison with results
obtained with the real materials discussed below.
\item Mirrors made of intrinsic silicon are described by
a Drude-Lorentz model \cite{Lambrecht2007,Messina2009}~:
\begin{equation}
\label{silicon} \epsilon(i\xi)=\epsilon_\infty
+\frac{(\epsilon_0-\epsilon_\infty)\omega_0^2}{\xi^2+\omega_0^2}~,
\end{equation}
with the parameters $\epsilon_0=11.87$, $\epsilon_\infty=1.035$,
$\omega_0=6.6\times 10^{15}$ rad.s$^{-1}$.
\item Mirrors made of amorphous silica are described by a simple
Sellmeier model \cite{springerhandbook}~:
\begin{equation}
\label{silica} \epsilon(i\xi)=1+\sum_{i=1,2,3}
\frac{B_i}{1+(\xi/\omega_i)^2}~,
\end{equation}
with the parameters $B_{1,2,3}=$ 0.696749, 0.408218, 0.890815 and
$\omega_{1,2,3}=$ 27.2732, 16.2858, 0.190257 $\times$
10$^{15}$rad.s$^{-1}$.
\item The electronic properties of graphene are described by a Dirac model leading to reflection
coefficients given in \cite{BordagPRB2009}.
\end{enumerate}

The potential \eqref{vcp} has well-known asymptotic behaviors at
short and long distances
\begin{equation}
\label{vlimits} V(z) \underset{z \ll \ell}{\to} -\frac{C_3}{z^3}
\quad,\quad V(z)\underset{z \gg \ell}{\to} -\frac{C_4}{z^4}~.
\end{equation}
where $\ell$ is a distance scale determined by the characteristic
atomic frequencies which enter the expressions of polarizability or
dielectric function. The short distance limit is identical to the
famous London/Van der Waals result while the long distance limit is
the so-called \emph{retarded} Casimir-Polder interaction which takes
into account that the finite speed of light comes into play at large
separations \cite{Casimir1946,Casimir1948}. The values given in Table \ref{coeffs} are obtained from the exact vdW/CP potential \eqref{vcp} and
given in atomic units.

\begin{table}[H]
\begin{center}
\begin{tabular}{|C{1cm} | C{1cm} | C{1cm} | C{1cm}|} 
\hline
 & {perfect} & {silicon}
& {silica}  \\ \hline
$C_3$ & 0.25 & 0.10 & 0.05 \\ \hline
$C_4$ & 73.6  & 50.3  & 28.1 \\ \hline
\end{tabular}
\caption{\label{coeffs}Coefficients $C_3$ and $C_4$ for the vdW/CP interaction for
$\Hb$ atoms above perfect mirrors, silicon and silica bulks~; the
values are given in atomic units $\Hart \Bohr^n$ for $C_n$ (Hartree energy $\Hart \simeq
4.3597$aJ~; Bohr radius $\Bohr \simeq 52.917$pm).}
\end{center}
\end{table}

\begin{figure}[th]
\centering
 \includegraphics[width=9cm]{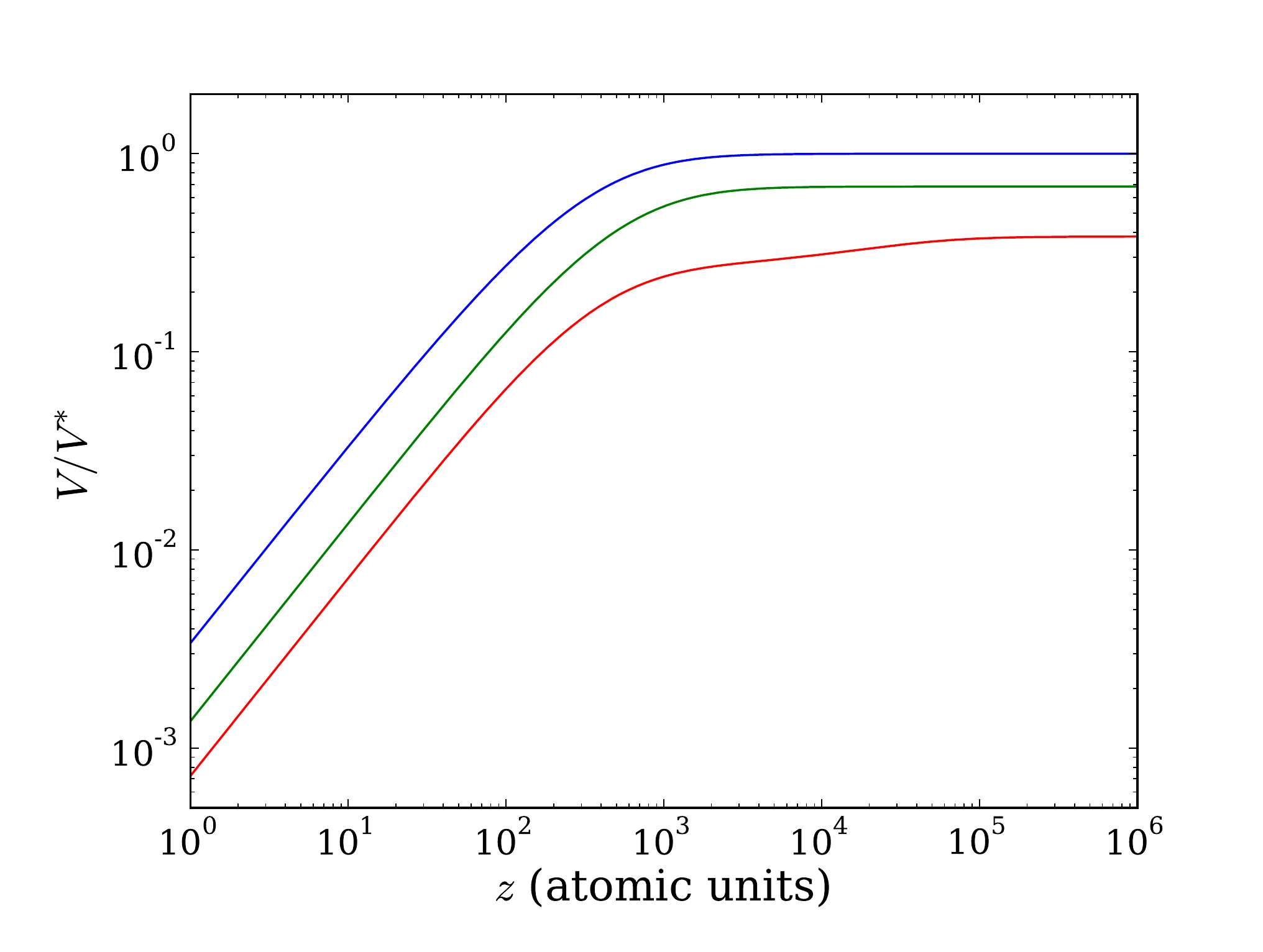}
\caption{(Color online) Casimir-Polder potential for $\Hb$ in the vicinity of a
material bulk, drawn as a ratio $V/V^*$ to the retarded ~potential
$V^*$ for a perfect mirror~; from top to bottom, perfect mirror
(blue), silicon (green), silica (red). }
\end{figure}

Figure 1 displays the exact vdW/CP potentials obtained from
\eqref{vcp} for $\Hb$ atoms on perfect mirrors and bulk mirrors made
of intrinsic silicon or amorphous silica, described by eqs.
\eqref{perfect}, \eqref{silicon} and \eqref{silica} respectively.
All cases are drawn as ratios $V(z)/V^*(z)$ to the retarded CP limit
calculated for a perfectly reflecting wall $V^*=-C_4^*/z^4$ with
$C_4^*=2.5 \cdot 10^{-57}$J m$^4$ (=73.6a.u., see Table \ref{coeffs}). The
ratios tend to constant values $C_4/C_4^*$ at large distances and
linear variations $C_3 z/C_4^*$ at small distances. Of course,
lesser and lesser reflective materials produce weaker and weaker CP
potentials, from perfect mirrors to silicon and silica plates.

\section{Quantum reflection of $\Hb$ }

We will now solve the problem of quantum reflection of $\Hb$ atoms
from the CP potential calculated in the previous section, starting
from free atoms with an energy $E>0$ just before they feel the CP
potential. We will also use below the notation $h$ for the height of
free fall of the atoms with the correspondence $E=mgh$ (supposing
$h$ much larger than tens of microns).

The Schr\"odinger equation may be written~:
\begin{equation}
\label{schrod} \psi^{\prime\prime}(z) +\frac{p^2(z)}{\hbar^2} \psi =
0~,
\end{equation}
where primes denote derivations with respect to $z$ while $p^2$ is
the square of the semiclassical momentum
\begin{equation}
p(z)=\sqrt{2m\left( E-V(z) \right)}~.
\end{equation}
The general solution can be expressed, without approximation, as a
superposition of the two WKB waves
\begin{eqnarray}
&&\psi \left( z\right) = \frac{c_+(z)}{\sqrt{\left\vert p(z)
\right\vert}} e^{i\phi\left( z\right)} +
\frac{c_-(z)}{\sqrt{\left\vert p(z) \right\vert}} e^{-i\phi\left(
z\right)} ~, \label{wkbwaves}
\end{eqnarray}
where $\phi$ is the WKB phase ($z_0$ arbitrary)
\begin{eqnarray}
&&\phi \left( z\right) = \int_{z_0}^{z} \frac{p (z^\prime) \dd
z^\prime}\hbar~.
\end{eqnarray}
The Schr\"odinger equation \eqref{schrod} is obeyed when the
amplitudes $c_\pm$ verify the coupled first-order equations
\cite{Berry1972}
\begin{equation}
c_\pm^{\prime}(z) = e^ {\mp 2i \phi \left( z\right) }
\,\frac{p^\prime(z)}{2 p(z)} \,c_\mp(z) ~.
\end{equation}

As $\Hb$ annihilates as soon as it touches the wall, there cannot be
any wave reflected immediately from the surface $z=0$ of the
material boundary \cite{Voronin2005,Voronin2005a}. This full
absorption condition imposes $c_+(z=0)=0$ and we are then free to
fix $c_-(z=0)=1$. The quantum reflection amplitude $r$ is thus given
by the ratio of the amplitudes $c_+(z)$ and $c_-(z)$ at the limit
$z\to\infty$ (see equation \ref{wkbwaves}). Finally, the quantum
reflection probability discussed below is the squared modulus of
this amplitude $\left\vert r\right\vert^2$.

In order to numerically integrate the preceding equations, it
remains to fix the problems arising from the divergence of the
potential in the vicinity of the surface. It will result from
forthcoming discussions that the WKB waves are well defined near the
wall. However a difficulty arises from the divergence of the WKB
phase $\phi$ there. To fix this difficulty, we proceed as in
\cite{Voronin2005, Voronin2005a} by studying the analytical form of
the solution for $c_\pm$ close enough to the wall. The potential
there takes its van der Waals approximated form while the energy $E$
becomes negligible when compared to the potential.

In this limit the functions $f_\pm(t)$ defined by 
$c_\pm(z)= x^{3/2} f_\pm(t)$, $x=\sqrt{8m C_3/z}$ and $t=\pm 2ix$
satisfy the Kummer equation:
\begin{equation}
t f''_\pm(t)+\left( b - t \right)f'_\pm(t) - a f_\pm(t)=0
\end{equation}
with parameters $a=3/2$ and $b=4$.
A pair of independent solutions is given by Kummer's confluent hypergeometric functions $M(a,b,t)$ and $U(a,b,t)$ \cite{kummer}.  
On the other hand the Schr\"odinger equation \eqref{schrod} can also be solved close to the wall.
The two counterpropagating waves can be expressed in terms of the Hankel functions as
$H^{(1)}_1(x)/x$ and $H^{(2)}_1(x)/x$ and the full absorption condition imposes that the second wave has a null amplitude
\cite{Voronin2005, Voronin2005a}.
By comparing this expression of the wave function with \eqref{wkbwaves} we find that
\begin{subequations}\label{wkb-coeffs-analytic}
\begin{align}
 c_+(x)=-2(1+i) x^{3/2} &\bigg[U\!\left( \frac{3}{2}, 4, 2ix \right) \\
&-\frac{i\sqrt{\pi}}{8} M\!\left( \frac{3}{2}, 4, 2ix \right) \bigg] e^{-2 i x_0} \notag\\
 c_-(x)=-2(1+i) x^{3/2} &U\!\left( \frac{3}{2}, 4, -2ix \right)
\end{align}
\end{subequations}

A better behavior of the functions is thus obtained by changing the
variables $z\to x$ in the vicinity of matter and matching the
numerical solutions to the known analytical solutions \eqref{wkb-coeffs-analytic}.
The results shown below are obtained in this manner close to the wall, while
variables are switched back to $z$ when the potential becomes
comparable to the energy.

\begin{figure}[thb]
\centering
 \includegraphics[width=9cm]{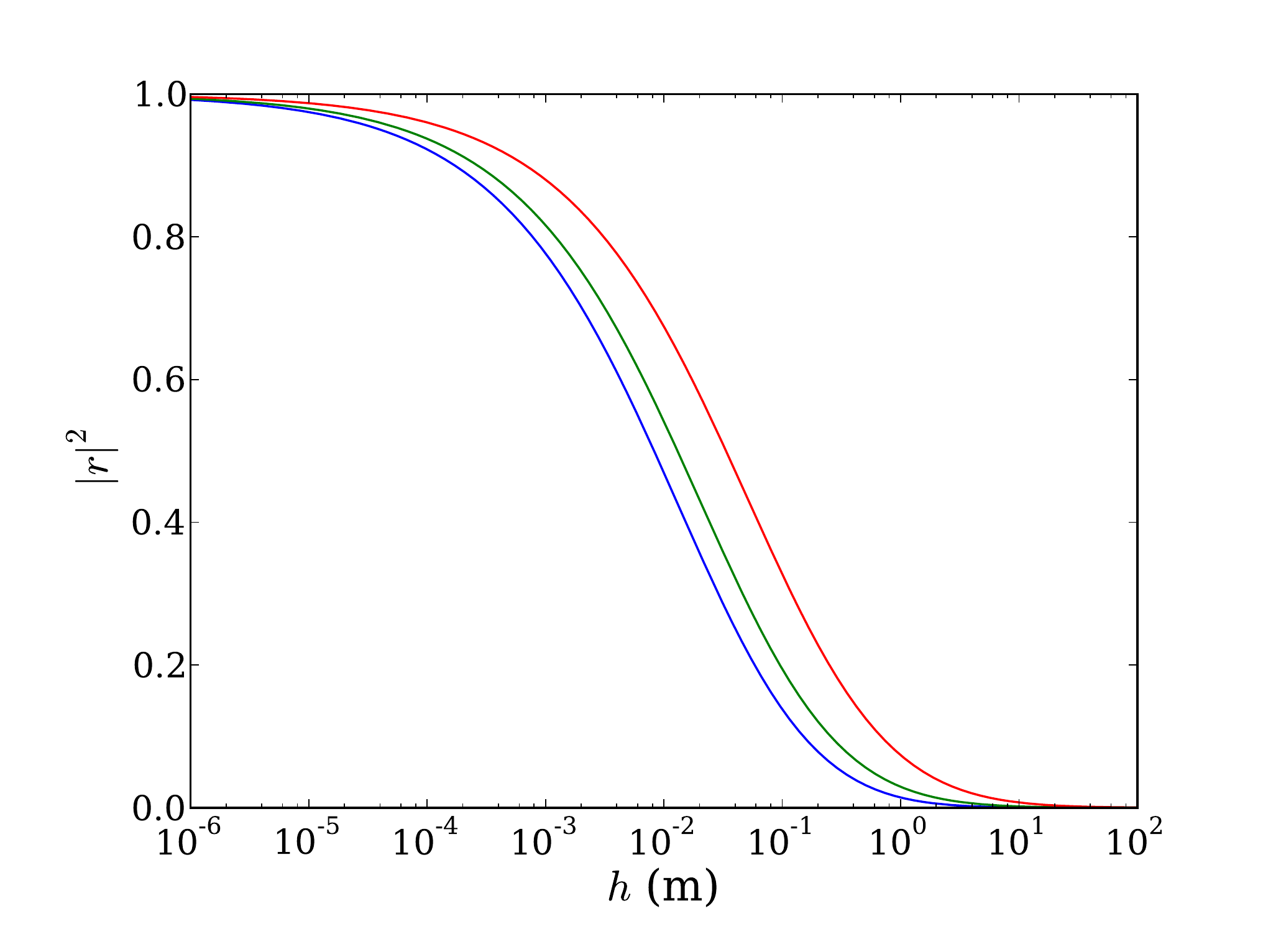}
 \caption{(Color online) Quantum reflection probability $\left\vert
r\right\vert^2$ as a function of the free fall height $h$ for $\Hb$
atoms on bulk mirrors~; from bottom to top, perfect mirror (blue),
silicon (green), silica (red). }
\end{figure}

We show in figure 2 the numerical solution for quantum reflection
probability obtained with the exact CP/vdW potentials discussed in
the preceding section for $\Hb$ atoms falling to perfect mirrors and
bulk mirrors made of silicon or silica. It turns out that
significant values are obtained for the quantum reflection
probability with the typical numbers considered for the project
GBAR as shown in Table \ref{reflectivities}.

\begin{table}[H]
\centering
 \begin{tabular}{|C{1cm} |C{1cm} | C{1cm} | C{1cm} m{0cm} |}
\hline
    &perfect  & silicon &  silica & \\
\hline
$\left\vert r\right\vert^2$& 14\% &20\% & 32\% & \\[1ex]
\hline
 \end{tabular}
\caption{\label{reflectivities} Reflection probabilities for a free fall height $h\sim10$cm, which
corresponds to an energy per unit mass $gh\sim1$(m/s)$^2$ at the
matter plate.}
\end{table}

These numbers highlight a striking result of the calculations which
is also emphasized by the use of the same color codes in figures 1
and 2~: when going to lesser and lesser reflective materials,
\textit{i.e.} weaker and weaker CP/vdW interactions, one indeed
obtains larger and larger quantum reflection probability
\cite{Judd2011,Pasquini2006}. This apparent paradox is analyzed in
the next section, by taking a closer look at the region where
quantum reflection occurs.

\section{Badlands condition}

We now discuss the so-called \emph{badlands} condition for efficient
reflection, that is also for significant non adiabatic transitions
beyond the WKB approximation \cite{Berry1972,Friedrich2004}.

To this aim, we recall that the WKB approximation $\tpsi$, the
wavefunction \eqref{wkbwaves} with constant amplitudes $c_\pm$, also
obeys a Schr\"odinger's equation~:
\begin{eqnarray}
\label{tschrod}  &&\tpsi^{\prime\prime}(z) +
\frac{\tp^2(z)}{\hbar^2} \tpsi(z) = 0~, \nonumber
\\
&& \tp^2(z) \equiv p^2(z) + \frac{\hbar^2}2 \calS\phi(z)~.
\end{eqnarray}
The difference between \eqref{schrod} and \eqref{tschrod} is the
extra term in $\tp^2$ with respect to $p^2$, proportional to the
Schwarzian derivative of the WKB phase~:
\begin{equation}
\calS\phi(z) \equiv
\frac{\phi^{\prime\prime\prime}(z)}{\phi^{\prime}(z)} - \frac32
\left( \frac{\phi^{\prime\prime}(z)}{\phi^{\prime}(z)} \right)^2~.
\end{equation}
This means that non adiabatic processes are characterized by this
Schwarzian derivative, in a similar way to non adiabatic emission of
photons in vacuum after reflection from moving mirrors
\cite{Fulling1976,Lambrecht1998}.

It follows that the WKB approximation is good when the second term
in $\tp^2$ in \eqref{tschrod} is much smaller than the first one. It
can be shown that this is the case for the problem being studied in
the present paper for short as well as long distances, which means
that left- and rightwards propagation are well defined in both
limits. The non adiabatic processes giving rise to quantum
reflection occur in the intermediate distance range, and their
efficiency is significant for large values of the quantity~:
\begin{equation}
Q(z) \equiv \frac{\hbar^2\calS\phi}{2p^2} = \frac{\hbar^2}{2}
\frac{p^{\prime\prime}(z)}{p(z)^3} - \frac{3 \hbar^2}{4}
\left(\frac{p^{\prime}(z)}{p(z)^2}\right)^2~.
\end{equation}
The adiabatic approximation breaks down in regions where $|Q(z)|
\sim 1$, which have been dubbed the \emph{badlands}. Non adiabatic
quantum reflection happens there, where the notions of left- and
rightwards propagation are no longer well defined.

\begin{figure}[thb]
\centering
 \includegraphics[width=9cm]{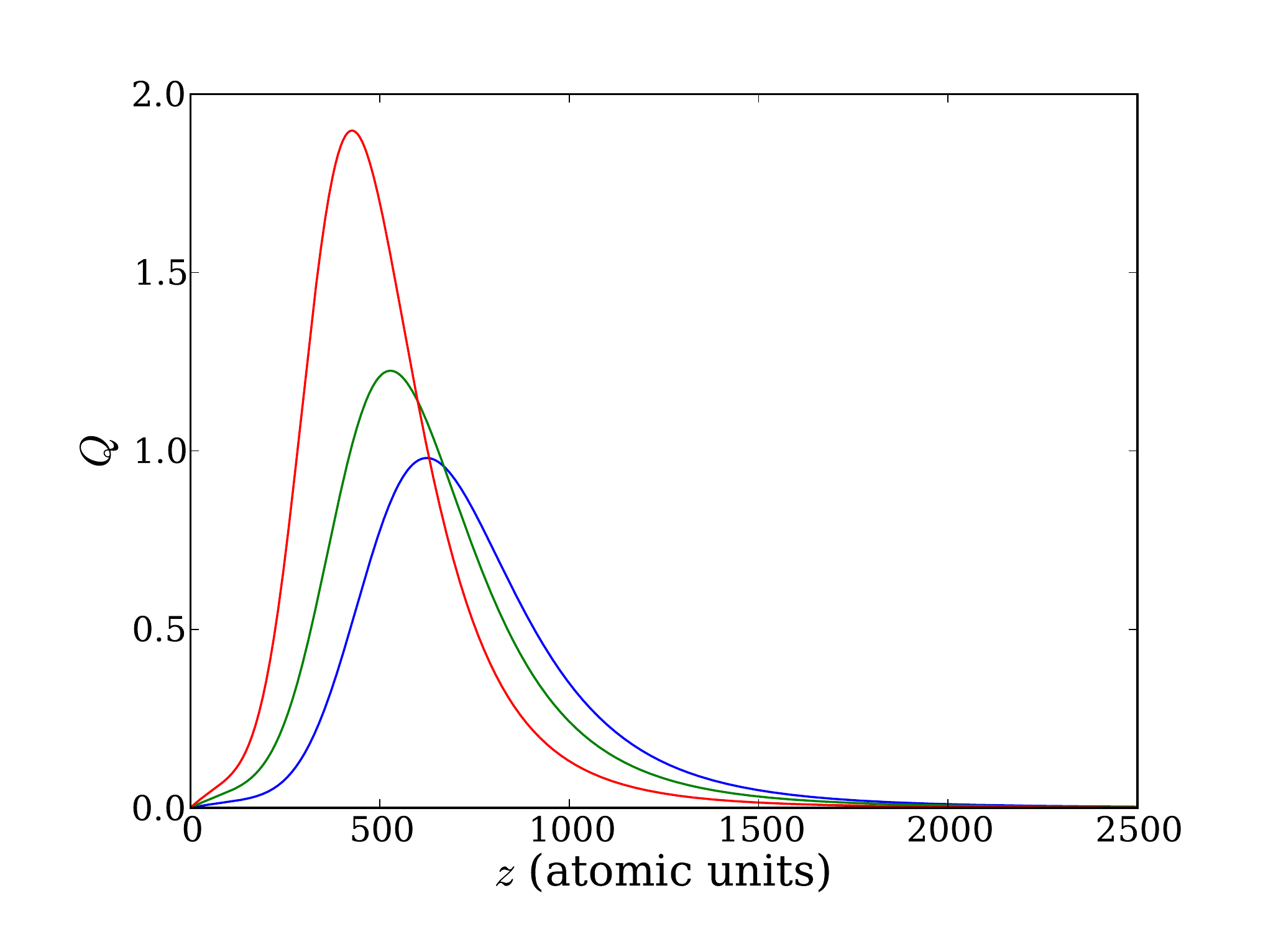}
\caption{(Color online) Badlands function $Q(z)$ as a function of distance to the
wall (atomic units) for $\Hb$ dropped from $h=10$cm on bulk
mirrors~; from bottom right to top left, perfect mirror (blue),
silicon (green), silica (red). }
\end{figure}

Figure 3 features the numerical evaluation of this badlands function
$Q(z)$ as a function of distance $z$ to the wall (atomic units), for
$\Hb$ dropped from $h=10$cm on perfect, silicon or silica mirrors
(same color codes as in figures 1-2). The plots clarify two features
which explain the apparent paradox discussed in the preceding
section. First, quantum reflection occurs closer and closer to the
wall for weaker and weaker CP/vdW interaction. Second, the value
reached by $Q(z)$ is thus larger and larger, since the CP
interaction gets steeper and steeper when atoms approach the wall.
When considered together, these two features explain why a weaker
potential leads to a more efficient reflection than a stronger one.
In fact, the quantum reflection probabilities $\left\vert
r\right\vert^2$ (see for example the numbers given in Table \ref{reflectivities}) increase with increasing peak
value of the badlands function $Q(z)$.

\section{Reflection on a thin slab}

This discussion suggests that one should try to weaken further the
CP/vdW interaction with the aim of enhancing quantum reflection
\cite{Judd2011}. In the present section, we analyze this idea by
studying either slabs having a finite thickness or a graphene layer.

The calculations proceed along the same lines as previously, except
for the fact that slabs of finite thickness $d$ have smaller
reflection amplitudes than the corresponding bulks. There is a
general relation between these amplitudes \cite{Lambrecht2007}~:
\begin{equation}
\label{r-slab}  \rho^p_\slab=\frac{\left(1-e^{-2\Kappa d}\right) \,
\rho^p_\bulk} {1-e^{-2\Kappa d}\,\left(\rho^p_\bulk\right)^2} ~.
\end{equation}
When the CP/vdW interaction is calculated at distances $z$ smaller
than the thickness $d$, the results of the bulk are recovered. This
can be understood from the fact that $\rho^p_\slab$ goes to
$\rho^p_\bulk$ for large values of $d$ (up to exponentially small
corrections). In contrast, the long-distance behavior of the CP
potential is completely changed because the exponential factor now
plays an important role in \eqref{r-slab}. Even the power law index
is changed for the potential which now varies as~:
\begin{equation}
\label{retarded-slab} V(z) \underset{z\gg \ell,d}{\to}
-\frac{C_5}{z^5}~.
\end{equation}

\begin{figure}[thb]
\centering
 \includegraphics[width=9cm]{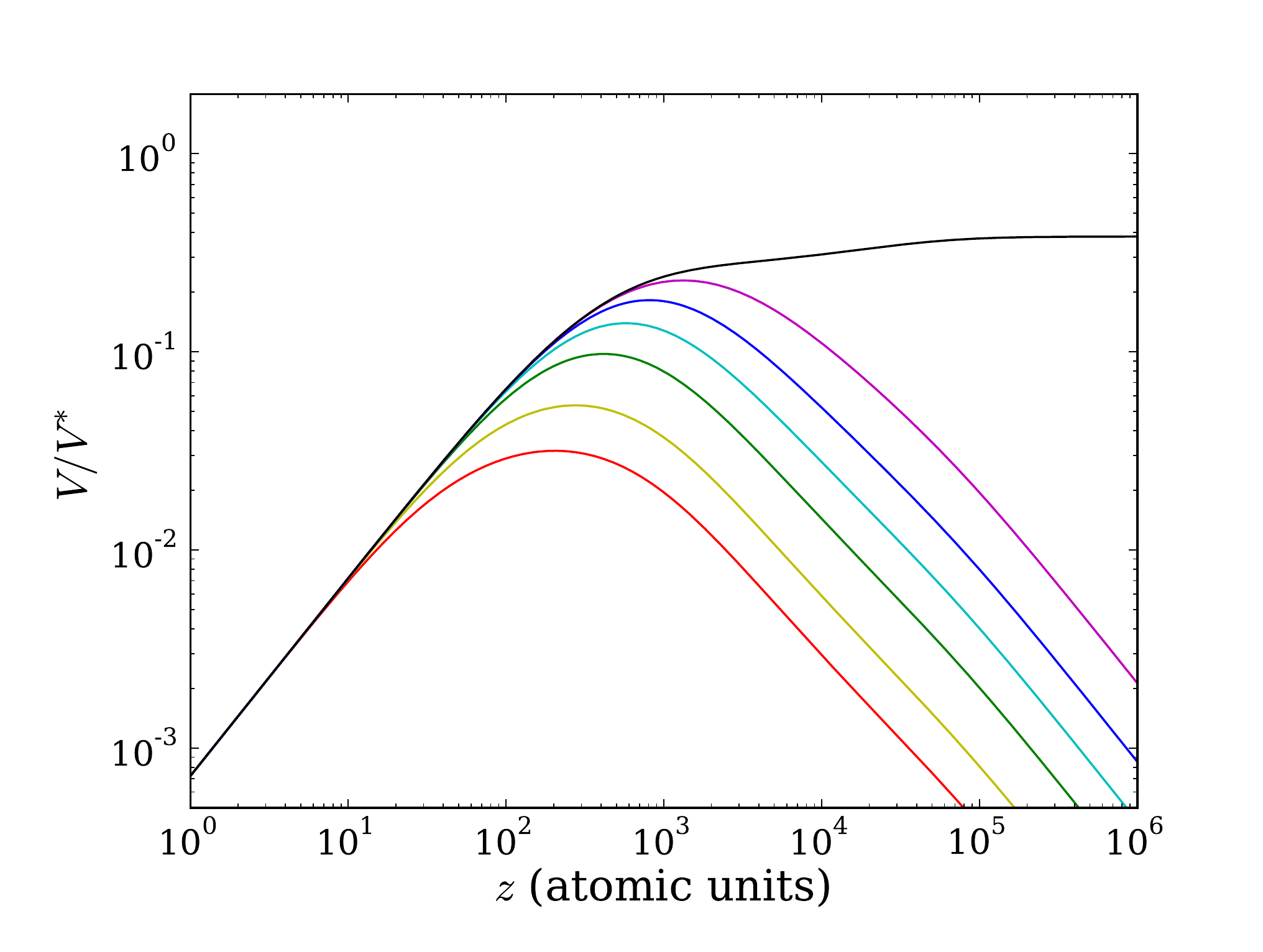}
\caption{(Color online) Casimir-Polder potential for $\Hb$ in the vicinity of a
silica slab, drawn as a ratio $V/V^*$~; from top to bottom, the
thickness is infinite (black), 50 nm (magenta), 20 nm (deep blue),
10 nm (light blue), 5 nm (deep green), 2 nm (light green) and 1 nm
(red). }
\end{figure}

Figure 4 shows the exact CP/vdW potentials obtained from \eqref{vcp}
for $\Hb$ atoms on slabs of amorphous silica, with different values
for the thickness $d$. All cases are drawn as ratios of $V(z)$ to
the same reference potential $V^*$ already used in figure 1. The
ratios tend to the same linear variations $C_3 z/C_4^*$ at small
distances as for the silica bulk (red curve in figure 1) and to
inverse distance laws $C_5/(C_4^*z)$ at large distances with the
value of $C_5$ being proportional to $d$. This behavior can be
expected from a simple argument where the potential $V_\slab(z)$ at
distance $z$ from a slab of thickness $d$ is obtained from the
difference $V_\bulk(z)-V_\bulk(z+d)$ with $V_\bulk$ the potential at
distance $z$ from a bulk. The scaling given by this simple argument
is correct while the value of $C_5$ is not exact.

\begin{figure}[thb]
\centering
 \includegraphics[width=9cm]{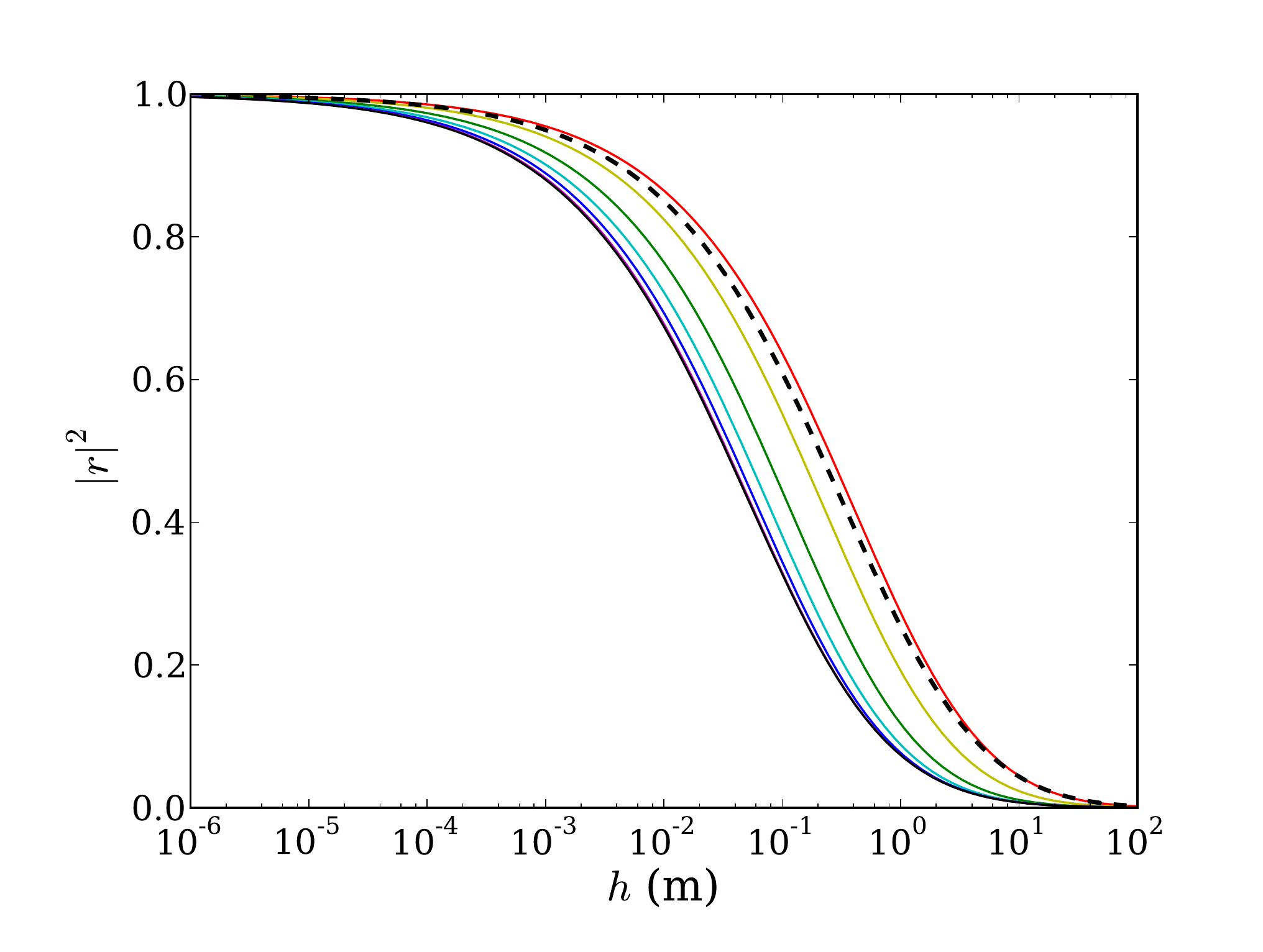}
 \caption{(Color online) Quantum reflection probability $\left\vert
r\right\vert^2$ as a function of the free fall height $h$ for $\Hb$
atoms on silica slabs~; from bottom to top, the thickness is
infinite (black), 50 nm (magenta), 20 nm (deep blue), 10 nm (light
blue), 5 nm (deep green), 2 nm (light green) and 1 nm (red). The
dashed line is the result for quantum reflection on non-doped
graphene. }
\end{figure}

We depict in figure 5 the numerical solution for quantum reflection
probability obtained with the exact CP/vdW potentials for $\Hb$
atoms falling to silica slabs with various values of the thickness
(same color code as on figure 4). As expected, larger and larger
values are obtained for the quantum reflection probability on
thinner and thinner silica slabs, that is also steeper and steeper
CP/vdW potentials. For a free fall height $h\sim10$ cm for example,
the probability $\left\vert r\right\vert^2$ reaches $\sim$ 50\% for
$3$ nm slabs while it is only 33\% pour Silica bulks. For comparison
we also show the quantum reflection coefficient for graphene.
Interestingly the same high quantum reflection than on a (not
realistic) 1 nm slab can be obtained with the quantum reflection
reaching 61\% for non-doped graphene. This value increases only
slightly ($\leq 2$\%) if doping is included.

\section{Low energy limit}

We finally discuss the limit of \emph{near threshold} quantum
reflection, where the incident atomic energy $E$ goes to zero.
Quantum reflection is thus characterized by a scattering length
\cite{Voronin2005,*Voronin2005a} which we will calculate in the
present section for the different cases discussed above, with the
aim of optimizing applications for manipulating antihydrogen with
material walls \cite{Voronin2011,Voronin2012,*Voronin2012a}.

In order to conform to standard notation, we replace $p$ by $\hbar
k$ in this section ($k$ not to be confused with the electromagnetic
wavevector used in the beginning of this paper). The reflection
amplitude $r$ is a function of $k$ which can be written in terms of
a complex-valued function $a(k)$ having the dimension of a length~:
\begin{equation}
r(k) = -\exp\left(-2 ik a(k)\right)~.
\end{equation}
The real part of $a(k)$ determines the phase at reflection while its
imaginary part determines the quantum reflection probability~:
\begin{equation}
|r|^2 = \exp\left(4 k \mathrm{Im}(a(k) \right) ~.
\end{equation}

\begin{figure}[h]
\centering
 \includegraphics[width=9cm]{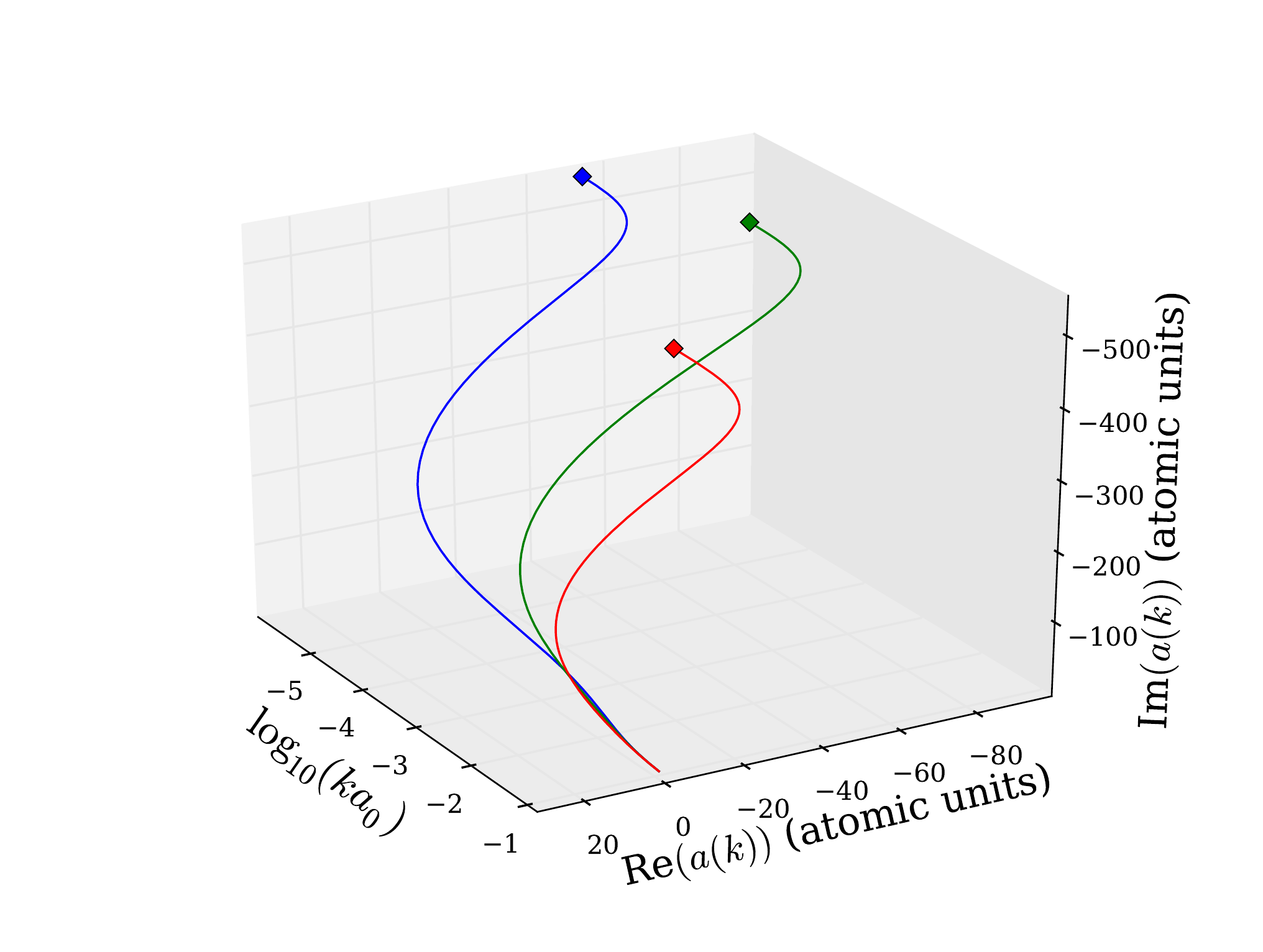}
\caption{(Color online) 3D representation of the variation of real and imaginary
parts of $a(k)$ versus wavevector for $\Hb$ atoms on bulk mirrors~;
from top left to bottom right, perfect mirror (blue), silicon
(green), silica (red). }
\end{figure}

We show in figure 6 the variations of real and imaginary parts of
$a(k)$ versus wavevector (measured in atomic units) for $\Hb$ atoms
falling to perfect mirror, silicon and silica bulks (same color
codes as in figure 2). We see that $a(k)$ goes to a finite value
$a(0)$ when $k\to 0$, which is known as the scattering length; the
values of $a(0)$ are collected in Table \ref{scatlength-mats}.

\begin{table}[H]
\centering
 \begin{tabular}{|c | c | c | c | c | c | c | c|}
\hline
\multicolumn{2}{|c |}{perfect} & \multicolumn{2}{c |}{silicon} & \multicolumn{2}{c |}{silica} &\multicolumn{2}{c|}{graphene}\\
\hline
Re($a$)  &Im($a$)   & Re($a$)  &Im($a$) &Re($a$) &Im($a$)&Re($a$) &Im($a$)  \\
\hline
-53.0&-543.0&-97.2&-435.2&-77.0&-272.6 &-15.4&-109.7\\
\hline
 \end{tabular}
\caption{\label{scatlength-mats}Real and imaginary parts of the scattering length $a(0)$
for $\Hb$ falling on perfect mirrors, silicon and silica bulks and
graphene (given in atomic units $\Bohr$).}
\end{table}

\begin{figure}[h]
\centering
 \includegraphics[width=9cm]{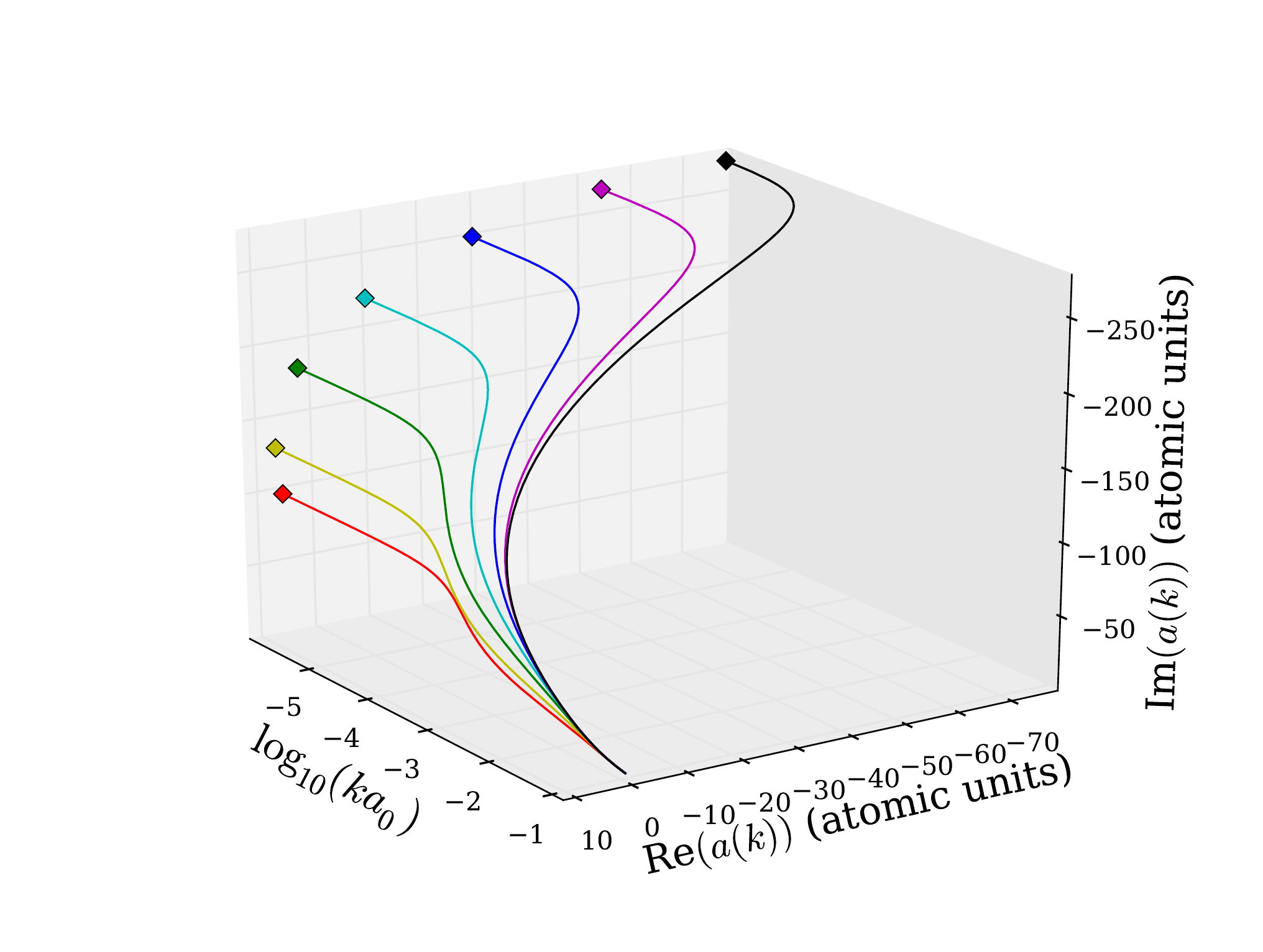}
\caption{(Color online) 3D representation of the variation of real and imaginary
parts of $a(k)$ versus wavevector $\Hb$ atoms falling to silica
slabs~; from top right to bottom left, the thickness is infinite
(black), 50 nm (magenta), 20 nm (deep blue), 10 nm (light blue), 5
nm (deep green), 2 nm (light green) and 1 nm (red). 
}
\end{figure}

We also show in figure 7 the variations of real and imaginary parts
of $a(k)$ versus wavevector (measured in atomic units) for $\Hb$
atoms on silica slabs (same color codes as in figure 5). Again,
$a(k)$ goes to a finite value $a(0)$ when $k\to 0$, the real and
imaginary parts of which are collected in Table \ref{scatlength-slabs}.

\begin{table}[H]
\centering
 \begin{tabular}{|c | c | c | c | c|}
\hline
$d$ (slab    & \multicolumn{2}{c |}{silicon} & \multicolumn{2}{c|}{silica} \\
         \cline{2-5}
 thickness)   & Re($a$)  &Im($a$) &Re($a$) &Im($a$)  \\
\hline
$1$nm&3.0&-178.1&6.5&-97.9\\
$2$nm&1.6&-231.8&7.5&-130.3\\
$5$nm&-6.5&-311.2&3.2&-181.9\\
$10$nm&-21.8&-367.8&-9.3&-221.1\\
$20$nm&-45.2&-408.0&-29.1&-250.1\\
$50$nm&-73.1&-429.7&-53.3&-267.4\\
$100$nm&-85.0&-433.7&-64.4&-271.2\\
bulk&-97.2&-435.2&-77.0&-272.6 \\
\hline
 \end{tabular}
\caption{\label{scatlength-slabs}Real and imaginary parts of the
scattering length of antihydrogen on silicon and silica slabs  (given in atomic
units $\Bohr$).}
\end{table}

We observe large variations of these values, which can have
important applications for manipulating $\Hb$ with material walls.
By considering quantum gravitational traps for $\Hb$ bounded below
by the quantum reflection from the CP/vdW potential and above by
gravity, one obtains the following lifetime for the quantum bouncer
in the first gravitational quantum state \cite{Voronin2011}~:
\begin{equation}
\tau=\frac{\hbar}{2 m g \left\vert\mathrm{Im}\,a(0)\right\vert}
\end{equation}
The lifetime is thus $\sim$ 5 times larger for thin silica slabs
than for the perfect mirrors considered in the calculations of
\cite{Voronin2011}. The same improvement holds for the width of
resonances between quantum states which can be used for precise
spectroscopic determination of the energies of these states, a
technique which could allow a better accuracy for the gravitational
behavior of $\Hb$ atoms in future experiments
\cite{Voronin2012,Voronin2012a}.

The same techniques could also allow to trap antiatoms above curved
material surfaces and them guide them at will during the longer
lifetime achieved thanks to quantum reflection from steep
potentials.

\section{Conclusion}

We have given realistic estimates of the VdW/CP potential above matter slabs of arbitrary thickness and the corresponding reflection probability for antihydrogen atoms. It appeared that a substantial amount of quantum reflection is to be expected in the GBAR experiment. We gave a detailed analysis of the reflection process, solving the paradox of weaker potentials leading to higher reflection. Finally we have investigated the low-energy regime of quantum reflection and given quantitative predictions for the scattering length.

\section{Acknowledgements}
The authors thank the ESF Research Networking
Programme CASIMIR (casimir-network.org) and the
GBAR collaboration (gbar.in2p3.fr) for providing
excellent possibilities for discussions and exchange.

\end{document}